\documentclass[12pt]{article}
\usepackage{times}
\usepackage{geometry}
\usepackage{floatflt}
\usepackage[utf8]{inputenc}
\usepackage{enumitem,amssymb}
\usepackage{ragged2e}
\newlist{thematic}{itemize}{8}
\setlist[thematic]{label=$\square$}
\usepackage{pifont}
\usepackage{jheppub}   
\usepackage[dvipsnames,svgnames,x11names]{xcolor}
\usepackage{sectsty}
\usepackage[outercaption]{sidecap}
\sidecaptionvpos{figure}{c}

\usepackage{caption}
\newcommand{\lya}{Ly$\alpha$}
\newcommand{\hone}{H~\textsc{i}}
\newcommand{\cfour}{C~\textsc{iv}}
\newcommand{\osix}{O~\textsc{vi}}
\newcommand{\hetwo}{He~\textsc{ii}}
\newcommand{\neeight}{Ne~\textsc{viii}}

\newcommand\hst{{\sl HST}}
\newcommand\hstcos{{\sl HST/COS}}
\newcommand\luvoir{{\sl LUVOIR}}

\newcommand\xrism{{\sl XRISM}}
\newcommand\athena{{\sl Athena}}
\newcommand\lynx{{\sl Lynx}}

\definecolor{DarkGreen}{rgb}{0.0, 0.3, 0.0}
\definecolor{purple}{rgb}{0.5, 0.0, 0.5}
\definecolor{red}{rgb}{1, 0.0, 0.0}
\definecolor{green}{rgb}{0, 1.0, 0.0}

\newcommand{\apj}{\rm ApJ}
\newcommand{\apjl}{\rm ApJL}
\newcommand{\apjs}{\rm ApJS}

\newcommand{\aap}{\rm A$\&$A}

\newcommand{\mnras}{\rm MNRAS}

\newcommand{\araa}{\rm ARAA}
\newcommand{\nat}{\rm Nature}

\newcommand{\procspie}{\rm Proc.\ SPIE}
\newcommand{\ssr}{Space Sci. Rev.}

\newcommand{\ndash}{$-$}
\newcommand{\lt}{\mbox{$<$}}

\def\3he{$^3{\rm He}$}

\def\lsim{\mathrel{\lower2.5pt\vbox{\lineskip=0pt\baselineskip=0pt
           \hbox{$<$}\hbox{$\sim$}}}}

\def\gsim{\mathrel{\lower2.5pt\vbox{\lineskip=0pt\baselineskip=0pt
           \hbox{$>$}\hbox{$\sim$}}}}

\usepackage{xcolor}
\usepackage{color}
\usepackage[normalem]{ulem}

\begin{document}
\raggedright
\uline{Astro2020 Science White Paper\hfill}\vspace{3mm}\linebreak
\huge
Ultraviolet Perspectives on Diffuse Gas in the Largest Cosmic Structures \linebreak
\normalsize

\noindent \textbf{Thematic Areas:} \hspace*{60pt} $\square$ Planetary Systems \hspace*{10pt} $\square$ Star and Planet Formation \hspace*{20pt}\linebreak
$\square$ Formation and Evolution of Compact Objects \hspace*{31pt} $\rlap{$\checkmark$}\square$ Cosmology and Fundamental Physics \linebreak
  $\square$  Stars and Stellar Evolution \hspace*{1pt} $\square$ Resolved Stellar Populations and their Environments \hspace*{40pt} \linebreak
  $\rlap{$\checkmark$}\square$    Galaxy Evolution   \hspace*{45pt} $\square$             Multi-Messenger Astronomy and Astrophysics \hspace*{65pt} \linebreak
  
\textbf{Principal Authors:}

Name: Joseph N. Burchett$^1$, Daisuke Nagai$^2$
 \linebreak						
Institution: 1) Univ. of California - Santa Cruz; 2) Yale University
 \linebreak
Email: burchett@ucolick.org; daisuke.nagai@yale.edu
 \linebreak
Phone:  +1 (831) 459-3081; +1 (203) 432-5370
 \linebreak
 
\textbf{Co-authors:} Iryna Butsky (UW-Seattle), Michael Tremmel (Yale), Rongmon Bordoloi (NC State), Greg Bryan (Columbia), Zheng Cai (UCSC), Rebecca Canning (Stanford), Hsiao-Wen Chen (U. Chicago), Alison Coil (UCSD), Drummond Fielding (Flatiron Institute), Michele Fumagalli (Durham), Sean D. Johnson (Princeton), Vikram Khaire (UCSB), Khee-Gan Lee (Kavli IPMU), Nicolas Lehner (U. Notre Dame), Nir Mandelker (Yale/Heidelberg), John O'Meara (Keck Observatory), Sowgat Muzahid (Leiden), Dylan Nelson (MPA), Benjamin D. Oppenheimer (CU-Boulder), Marc Postman (STScI), Molly S.\ Peeples (STScI/JHU), Thomas Quinn (UW-Seattle), Marc Rafelski (STScI/JHU), Joseph Ribaudo (Utica College), Kate Rubin (San Diego State), Jonathan Stern (Northwestern), Nicolas Tejos (PUCV), Stephanie Tonnesen (Flatiron Institute), Todd Tripp  (UMass-Amherst), Q. Daniel Wang (UMass-Amherst), Christopher N. A. Willmer (Steward Observatory), Yong Zheng (UC Berkeley)

\justify
\textbf{Abstract:} The past decade has seen an explosion of discoveries and new insights into the diffuse gas within galaxies, galaxy clusters, and the filaments composing the Cosmic Web. A new decade will bring fresh opportunities to further this progress towards developing a comprehensive view of the composition, thermal state, and physical processes of diffuse gas in the Universe.  Ultraviolet (UV) spectroscopy, probing diffuse $10^4-10^6$ K gas at high spectral resolution, is uniquely poised to (1) witness environmental galaxy quenching processes in action, such as strangulation and tidal- and ram-pressure stripping, (2) directly account for the baryon content of galaxy clusters in the cold-warm ($T<10^6$K) gas, (3) determine the phase structure and kinematics of gas participating in the equilibrium-regulating exchange of energy at the cores of galaxy clusters, and (4) map cold streams and filaments of the Cosmic Web that feed galaxies and clusters. With a substantial UV undertaking beyond the Hubble Space Telescope, all of the above would be achievable \textit{over the entire epoch of galaxy cluster formation}.  Such capabilities, coupled with already-planned advancements at other wavelengths, will transform extragalactic astronomy by revealing the dominant formation and growth mechanisms of gaseous halos over the mass spectrum, settling the debate between early- and late-time metal enrichment scenarios, and revealing how the ecosystems in which galaxies reside ultimately facilitate their demise.

\clearpage
\section{UV Frontiers: The CGM to Galaxy Clusters \& Cosmic Web}
\vspace{-2mm}
Enormous progress has been made over the last decade in our understanding of the diffuse gas within the circumgalactic medium (CGM), intracluster medium (ICM), and intergalactic medium (IGM) in the Cosmic Web.  Ultraviolet (UV) spectroscopy has been the primary driving force behind advancements in the CGM, while X-ray and radio techniques have predominantly been employed for groups and clusters of galaxies.  A new decade brings fresh opportunities to build on this multiwavelength progress towards unraveling the composition, thermal state, and physical processes within the most massive structures in the Universe, which bear directly on galaxy evolution, structure formation, and cosmology.  

The CGM is a critical piece of the ecosystems within which galaxies live, breathe, and die (see White Paper by Peeples et al.). We have seen a progression from detecting/confirming/characterizing the presence and composition of the CGM \cite{Bergeron:1991qy,Tripp:1998kq,Chen:2001ys,Stocke:2006yu,Prochaska:2011aa} to leveraging diagnostics from larger datasets and informing rigorous theoretical pursuit of the intimate connection between galaxy evolution and the CGM \cite{Tumlinson:2017aa,McQuinn:2018aa,Faerman:2017aa,Werk:2016aa,Oppenheimer:2016lr,Nelson:2018aa,Nielsen:2016aa}.  Among the notable CGM discoveries are (1) the CGM around star-forming galaxies is abundant in the gas traced by \osix\ while the CGM of quiescent galaxies is deficient \cite{Tumlinson:2011kx,Johnson:2015qv}, (2) the cool and warm-hot phases of the CGM potentially comprise enough mass to solve the `missing baryons' problem on galaxy scales \cite[for L* galaxies;][]{Werk:2014kx,Prochaska:2017aa}, and (3) the cool gas contents of the CGM are highly dependent on the galaxy environment \cite{Johnson:2014rt, Johnson:2015qv,Burchett:2016aa,Burchett:2018aa}.  These advances have all come through UV absorption line spectroscopy of background QSOs. Particularly aided by the sensitivity of the Cosmic Origins Spectrograph (COS) aboard \textit{Hubble Space Telescope} (\hst), we are now able to design absorption line experiments focusing on particular classes of galaxies, e.g., L* galaxies \cite{Tumlinson:2013cr}, dwarfs \cite{Bordoloi:2014lr,Burchett:2016aa,Johnson:2017aa}, and luminous red galaxies \cite{Chen:2018aa,Smailagic:2018aa, Berg:2018aa}. 

UV astronomy is poised to bring a unique but critical perspective to diffuse gas physics, from galaxies to galaxy clusters and the Cosmic Web, through the combination of (a) exclusive access to spectral transitions from cool ($10^4$ K) to warm ($10^5-10^6$ K) gas and (b) unrivaled spectral resolution capability for the physical processes of interest. Some progress has been made to apply similar approaches to more massive structures, such as galaxy clusters and large scale filaments and voids \cite[e.g.,][]{Yoon:2012yu,Burchett:2018aa, Muzahid:2017lr, Tejos:2012lr, Tejos:2016qv}, but this body of work is decidedly much less mature.  Progress is partly hindered by the fact that massive halos are rarer than less massive halos. This scarcity, coupled with the underlying paucity of viable UV-bright background sources such as QSOs, have limited the feasibility of building large statistical samples.

Although focused efforts with \hstcos\ can make great strides in setting benchmarks for cosmological hydrodynamical models, more advanced space-borne UV-sensitive assets, such as the Large Ultraviolet Optical Infrared \cite[\luvoir;][]{Bolcar:2017aa} observatory, stand to bring about a revolution in our understanding of gas flows, enrichment, and ultimately galaxy evolution on the largest scales.
As currently planned, \luvoir's 15m aperture will provide a factor of 50 increase in collecting area over \hst, and improvements in detector and mirror coating technology will boost throughput dramatically and to broader wavelength coverage.  In addition, multi-object spectroscopy \citep{Harris:2018aa} via a micro-shutter array will provide integral field spectroscopy over a 3'$\times$3' field of view.  The monumental increase in sensitivity provided by \luvoir\ translates into two key practical observational implications: (1) the number density of background sources feasibly observed for absorption line spectroscopy increases by $>3$ orders of magnitude and (2) sources for which we can readily obtain signal-to-noise ratio (S/N) of $\sim10$ with \hstcos\ may yield S/N $ >50$ with similar integration times.  In this White Paper, we highlight key science cases where UV spectroscopy will provide unique insights into the most massive structures in the Universe, and we discuss how current (\hstcos) and future (\luvoir) missions can deliver transformative understanding of galaxy evolution, galaxy cluster physics, and gas within the Cosmic Web. 

\begin{figure}[t]
    \centering
    \includegraphics[height=0.29\textwidth]{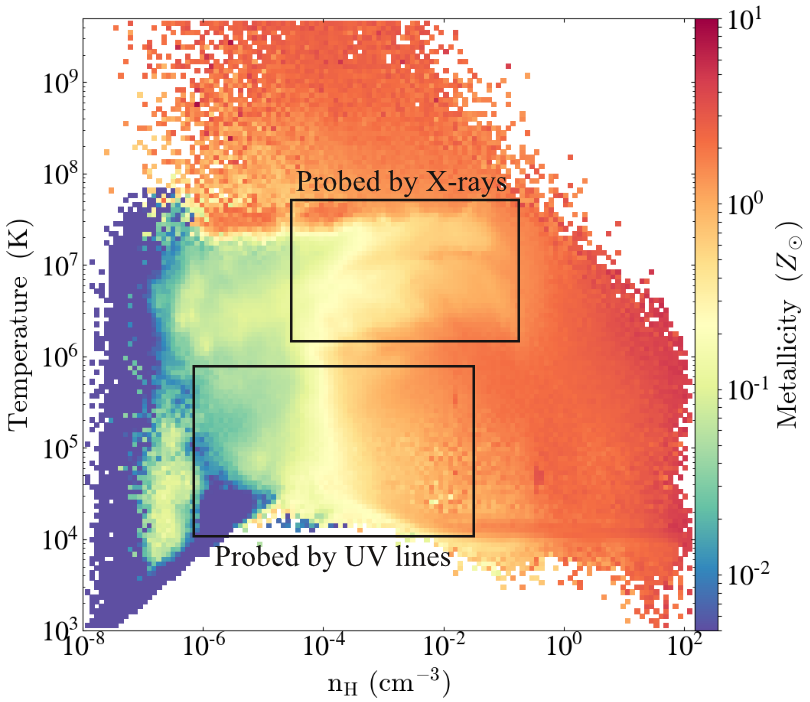}
    \includegraphics[height=0.29\textwidth]{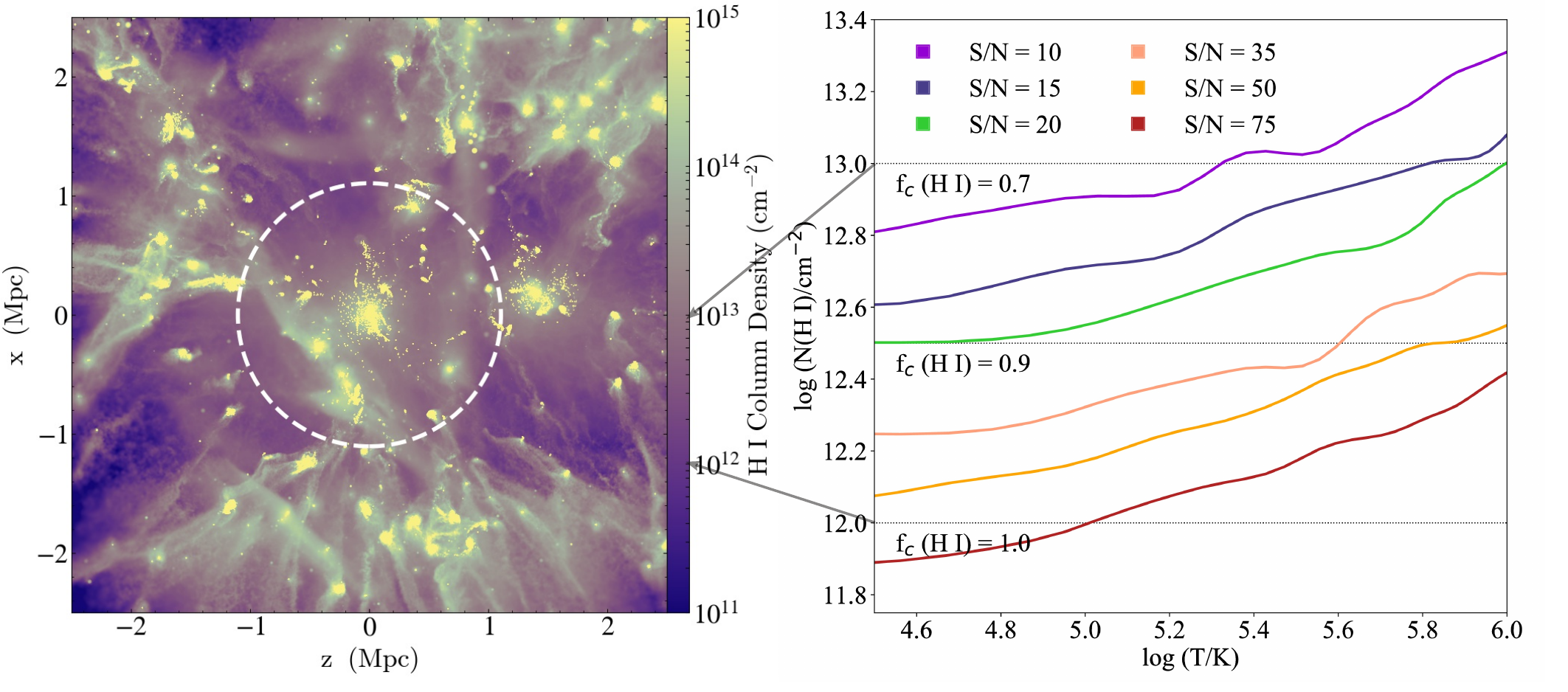}
    \caption{Properties of the cold-warm gas in the RomulusC high-resolution galaxy cluster simulation, with a mass of $M_{\rm 200} = 10^{14} M_\odot$ at redshift $z=0.3$. \textit{Left:} The mass-weighted metallicity in bins of temperature and density. Indicated are the regimes of phase space for which X-ray and UV observations are sensitive.  While the gas probed by X-rays is of nearly uniform enrichment, that probed by the UV exhibits a much wider range of metallicity. \textit{Center:} A projection of the predicted H I column density distribution (integration depth 5 Mpc).  The dashed white circle denotes R$_{200}$. Diffuse gas readily probed by UV spectroscopy is pervasive throughout the cluster, particularly towards the outskirts.  \textit{Right:} Column density detection limits of UV H I Lyman-$\alpha$ lines as a function of temperature assuming the lines are thermally broadened.  The curves correspond to different S/N levels ranging from those often obtained with \hstcos\ (S/N$\sim$10-15) to those readily achievable with an advanced UV mission such as the 15-m \luvoir\ (S/N$\sim$50-75).  Covering fraction predictions from RomulusC corresponding to 3 column density limits are marked with horizontal dotted lines on the right; two are connected to their corresponding shading in the map on the left. The next generation of UV telescopes will use H I in
    absorption to uncover the galaxy cluster structures invisible to X-rays. \vspace{-2mm}
     }
    \label{fig:romH1}
    \vspace{-3mm}
\end{figure}

\vspace{-4mm}
\section{Galaxy Clusters: a new frontier at all wavelengths}
\vspace{-2mm}

\noindent
{\bf CGM stripping and chemical enrichment in galaxy clusters:} Galaxy clusters form at the nodes of the cosmic web and are the densest pockets of the Universe. 
Recent multiwavelength observations (ranging from microwave to optical and X-ray) of galaxy clusters provide unprecedented views of the distribution of dark matter, gas and stars, enabling a plethora of new insights into the physics of both cluster cores \cite[e.g.,][]{McNamara:2005aa} and outskirts \cite{Walker:2019aa}.  The outskirts of galaxy clusters mark an exciting new territory for understanding how the clusters connect to the cosmic web, and they offer a powerful laboratory for studying the properties of the X-ray emitting ICM, chemical enrichment processes of the ICM, and evolution of galaxies in dense environments.  However, the cold-warm gas in cluster outskirts and around infalling galaxies remains elusive and largely unexplored.

Modern cosmological simulations predict that the relative fraction of $10^{4-6}$K gas greatly increases beyond the cluster virial radius \cite[Butsky et al., in prep;][]{Emerick:2015aa}, as also expected given evidence for a shock at $\sim$R$_{\rm vir}$ in SZ data \cite{Hurier:2019aa}.  UV absorption line surveys of cluster outskirts could discern between competing models, which vary in predicting how quickly these cool/warm gas fractions rise and how far into the outskirts they begin to exceed the hot gas.  The cold-warm gas properties in cluster outskirts are especially important, because they contain crucial information about how the metal-rich CGM of infalling cluster galaxies are stripped and subsequently pollute the chemical content of the ICM \citep{Gunn:1972qy,Fumagalli:2014aa, Jachym2014, Tonnesen:2007yq, Zinger:2018aa, Cramer2019}. As such, further studies of the cold-warm gas in galaxy clusters promise new insights into the following: {\it Where and how is the CGM of infalling galaxies stripped through interactions with the ICM? What quenching mechanisms are most important in high density environments? How do metals spread in the ICM? What is the role of feedback on the thermodynamic and chemical properties of the CGM and IGM?}

UV spectroscopy can bring a novel perspective to the cold-warm gas in galaxy clusters. Figure \ref{fig:romH1} shows an H I column density map from the high resolution galaxy cluster simulation RomulusC \cite[Butsky et al., in prep;][]{Tremmel:2019aa}.  H I is clearly abundant throughout the cluster at column densities that are near the detection limits for S/N$\sim$10 spectra, which are relative routine for \hstcos\ observing QSOs with m$_{\rm FUV} \lesssim  19$.  At these brightnesses, one can feasibly construct samples of background QSO/foreground cluster pairs.  Indeed, the few studies targeting clusters with QSO sightlines generally show a comparable detection rate of H I \cite{Yoon:2012yu, Yoon:2017aa, Burchett:2018aa}. However, there appears to be a dearth of H I absorbers at small velocity separation from the cluster redshift and at very small impact parameters, suggesting that the gas in the very inner regions is more highly ionized. 

Early observational evidence indicates stark differences between the CGM of cluster and field galaxies.  
For example, the CGM of cluster galaxies are highly depleted, with an \hone\ covering fraction of 25\% versus  nearly $100\%$ for field galaxies \cite{Burchett:2018aa}, illustrating increasing environmental influence on the composition, kinematics, and ionization state of the CGM \cite{Wakker:2009fr, Yoon:2013kq, Burchett:2016aa, Pointon:2017aa, Nielsen:2018aa}. A large sample of QSO sightlines probing clusters, coupled with follow-up galaxy spectroscopy, can make a good deal of progress in determining where upon infall and to what degree galaxies are stripped. Such experiments will also inform how the stripping of cluster galaxies contributes to the multiphase structure and metal content of the ICM on all scales.

Beyond their high sensitivity to the diffuse gas, another huge advantage of UV techniques to this field is their {\it high spectral resolution}.  
The highest resolution modes of COS reach FWHM $\sim 18$~km/s.  
With the resolution achievable in the UV, and given sufficient S/N, individual low-column density cool clouds (with narrow line profiles) will be easily distinguishable from warmer clouds with broad profiles. 
UV constraints on the kinematic properties of stripped CGM in galaxy clusters will be highly complementary to the bulk and turbulent gas motions of the hot ICM, which will be provided by ongoing and upcoming high-resolution X-ray (e.g., \xrism, \athena, \lynx) and SZ spectral imaging observatories (e.g., \textit{CCAT-prime, NIKA2, MUSTANG2, TolTEC, AtLAST, LST, CSST, CMB-in-HD}) in the coming decade \cite[][for recent reviews]{Simionescu2019,Mroczkowski2019}. 
\begin{figure}[t]
    \centering
    \includegraphics[width=\textwidth]{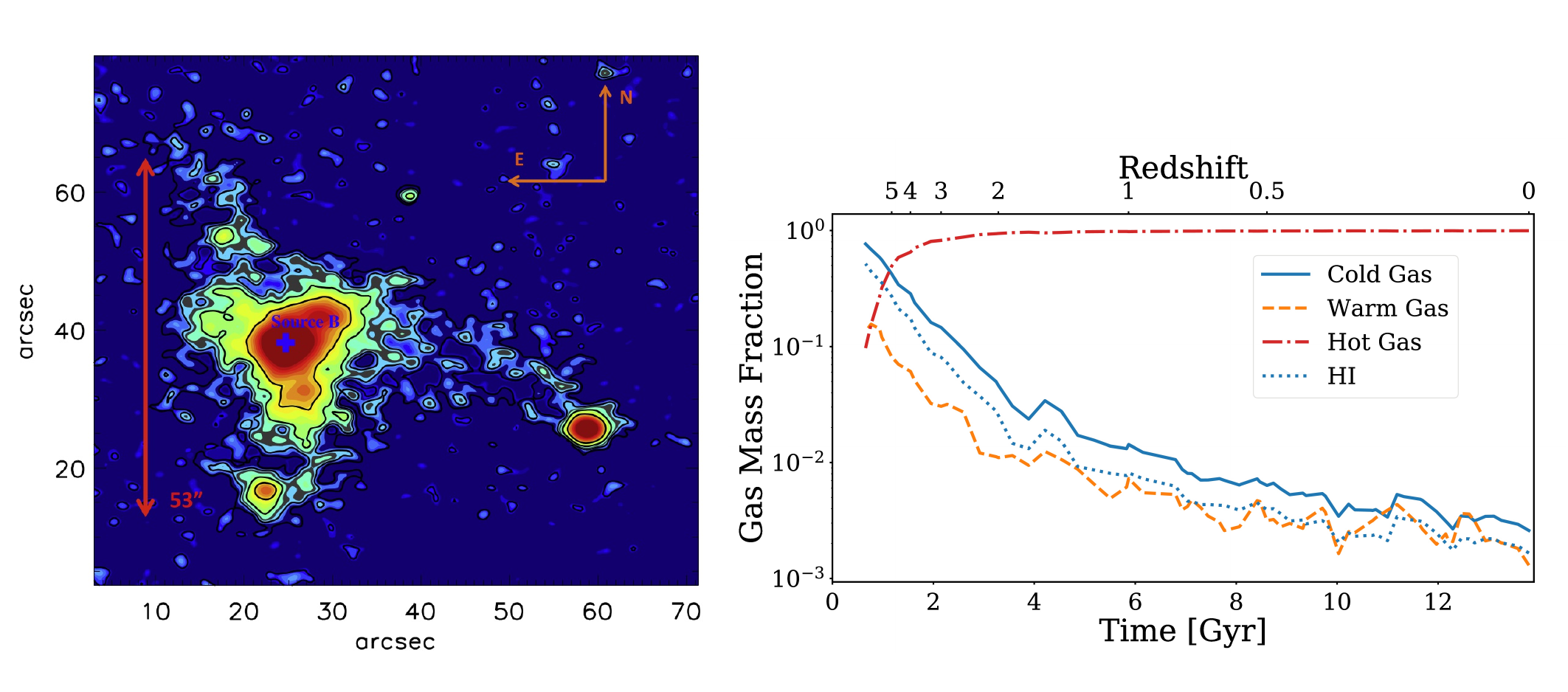}
    \vspace{-8mm}
    \caption{\textit{Left:} A $>400$ kpc enormous \lya\ nebula discovered witin a $z=2.3$ protocluster \cite{Cai:2017aa}, revealing a large reservoir of cold gas in the early stages of cluster formation. \textit{Right:} The redshift evolution of $10^4$ (cold), $10^{5-6}$~(warm), and $>10^6$ K (hot) gas mass fractions within R$_{200}$ of the RomulusC simulated galaxy cluster. The temperature distribution in clusters
    changes dramatically with redshift at $z\gtrsim2$. Systematically pursuing observations of Ly$\alpha$ and metal line-emitting nebulae (left panel) will place rigid constraints on the evolution of the ICM and cluster members. \vspace{-2mm}}
    \label{fig:elanEvo}
\end{figure}

\medskip
\noindent
{\bf Formation and evolution of cluster cores over cosmic time:}
Progress has also begun in quantifying the cold gas contents of clusters in their infancy. The left-panel of Figure \ref{fig:elanEvo} shows a $>400$ kpc \lya\ nebula, which also exhibits extended \cfour\ and \hetwo\ emission, in a $z = 2.3$ protocluster discovered using narrowband imaging and slit spectroscopy \cite{Cai:2017aa}. A large $>100$ kpc \lya\ nebulae in the core of an X-ray emitting galaxy cluster at $z = 1.99$ has also been detected \cite{Valentino:2016aa}.  
The presence of such material, particularly in the core of such a massive virialized halo (and observed on smaller scales at $z\sim0$ \cite{ODea:2004aa,Fabian:1984aa}), poses important questions as to its origins: {\it Are streams of gas readily able to penetrate deep into these massive halos, potentially providing fuel for star formation in the resident galaxies at high redshift \cite{Zinger2016,Mandelker:2019aa}?  Are we witnessing condensation directly out of the hot cluster atmosphere at early times, perhaps taking part in a self-regulating feedback process that feeds AGN activity and in turn injects energy into the surrounding CGM and ICM \cite{Voit:2015aa}?}  

These recent discoveries described above point towards a broader opportunity to track the evolution of galaxy clusters from the early protocluster phase through the mature ecosystems we observe at present times.  
Figure \ref{fig:elanEvo} (right) shows one prediction of the evolving mass fraction of $10^4$, $10^{5-6}$, and $>10^6$ K gas in a simulated cluster.  While much is to be learned at $z>2$, empirically constraining this evolution to any later times using ground-based instrumentation has already run into the unforgivingly hard wall of the UV atmospheric cutoff.  A space-based observatory with integral field spectroscopic capability, such as the Large UV Multi-Object Spectrograph \cite[LUMOS;][]{Harris:2018aa} micro-shutter array aboard \luvoir, or a balloon-borne experiment like FIREBall \cite{Tuttle:2008aa} could image the UV line-emitting diffuse gas \textit{and measure the kinematics} within clusters all the way to $z\sim2$, where telescopes on the ground can take over. The cool-warm gas constraints provided by rest-frame UV transitions, such as \lya, \cfour, and \hetwo\ already observed at $z\gtrsim2$, will provide benchmarks across cosmic time for cluster formation models. 

\begin{figure}[t]
    \centering
    \includegraphics[width=0.85\textwidth]{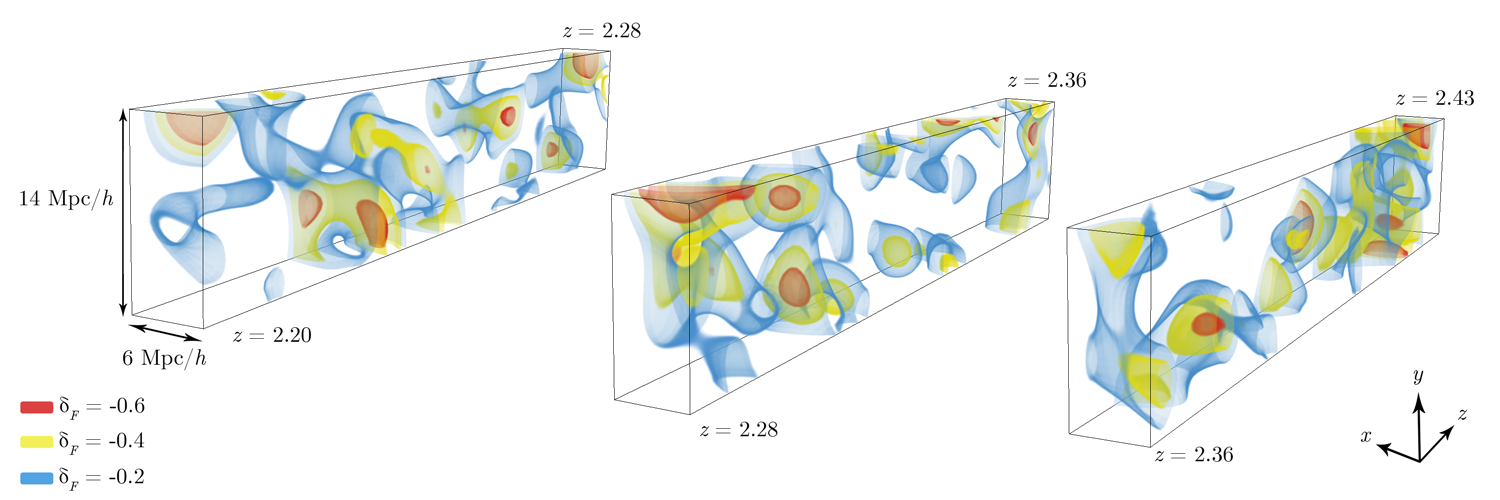}
    \vspace{-2mm}
    \caption{A tomographic map of filaments and voids in the Cosmic Web reconstructed from \lya\ absorption at $z>2$ against background galaxies \cite{Lee:2016aa}.  Colors represent \lya\ transmission, with red regions corresponding to the most overdense regions inferred from \lya\ absorption. The CLAMATO team achieved a high density of sightlines through this volume by leveraging relatively faint galaxies as background sources rather than QSOs.  Even higher sightline density may be achieved down to $z=0$ by coupling this technique with a large aperture UV facility. \vspace{-2mm}}
    \label{fig:tomography}
\end{figure}

\vspace{-4mm}
\section{The Cosmic Web}
\vspace{-2mm}

On cosmic, several novel methods have been employed to attempt mapping gas in the filaments, sheets, and voids composing the Cosmic Web, including stacking the SZ effect signal between massive halos \cite{Tanimura:2019aa, de-Graaff:2017aa},  \lya\ absorber statistics \cite{Tejos:2015lr,Tejos:2012lr}, and \lya\ forest tomography \cite{Cai:2016aa,Lee:2018aa}.  We focus on this last method to highlight prospects for studying the Cosmic Web given potential upcoming UV capability.  
Figure \ref{fig:tomography} shows a reconstructed map of Cosmic Web structure traced by \lya\ forest absorption in ground-based spectra of background star forming galaxies.  By using background galaxies instead of quasars, the CLAMATO project increased the projected sightline density from 80 deg$^{-2}$ to 1500 deg$^{-2}$.  Such sightline densities would be possible for $z<2$ (recall the hard redshift limit for ground-based \lya\ surveys) under the current \luvoir\ specifications, with the added efficiency of multi-object spectroscopy for simultaneously observing multiple sightlines.

Lastly, filaments in the Cosmic Web have been of extremely high interest due to their purportedly housing the bulk of "missing baryons" in the form of warm-hot ($10^{5-6}$K) intergalactic medium \cite[WHIM; e.g.,][]{Fukugita:1998vn,Dave:2001dp,Bregman2007}. In addition to broad \lya\ features, the extreme UV provides a relatively robust tracer of the WHIM in the \neeight\ 770, 780 \AA\ doublet.  The precipitous decline in \hst's sensitivity below 1150 \AA\ renders \neeight\ features effectively unreachable at $z<0.5$. Furthermore, \neeight\ features that trace the low density WHIM are expected to be very weak \cite{Tepper-Garcia:2013aa}. Herein lies a prime opportunity for future space-based missions: by providing decent throughput below 1000 \AA, they enable galaxy surveys sufficiently wide and deep to map out the large scale galaxy distribution and, e.g., separate circumgalactic \neeight\ \cite{Burchett:2018aa} from truly intergalactic material.

\vspace{-4mm}
\section{Prospects for the Next Decade}
\vspace{-2mm}
Here, we summarize the resources that can be leveraged now with \hstcos\ and in the future with the \luvoir\ observatory as currently conceived.  

\medskip
\noindent
\textbf{\hstcos}: Large surveys of cluster/QSO sightline pairs can provide a census of the cool-warm gas contents of galaxies, their outskirts, and pre-accretion shock region.  
High spectral resolution $\delta v \sim 20$ km/s enables kinematic separation of physically distinct cool, narrow-line absorption components and warm, broad-line components, which can in turn help identify bulk flows and kinematically connect absorbers to galaxies undergoing gas stripping within the cluster.

\smallskip
\noindent
\textbf{\luvoir}: A factor of $>50$ in sensitivity over \hst\ means the ability to (1) obtain extremely high S/N ($>$50) in the same amount of time for the same sources we observe now with \hst\ and (2) feasibly observe $>1000$ background sources per square degree on the sky.  Assuming a cluster with $M_{\rm 200} = 10^{14} M_\odot$ at $z=0.3$, this source density translates to $\gtrsim$16 potential background sources for \textit{any} cluster with at least this mass being observable for absorption studies. The increased sensitivity plus wavelength coverage down to 1000 \AA\ will provide a full suite of metal line diagnostics from low-, mid-, and high-ionization species to enable detailed modeling of the physical conditions of gas in any environment.  Integral field spectroscopic capability will enable imaging the diffuse gas emission \cite{Corlies:2016aa, Corlies:2018aa}, e.g., resolving its geometry and kinematics.

\medskip
\noindent
UV spectroscopy provides unique insights into the cold-warm gas in and around most massive structures in the Universe, providing highly complementary views of the baryonic contents of the universe provided by X-ray and microwave observations (see white papers on these topics).
When taken together, these forthcoming multiwavelength observations will provide a comprehensive view of the gaseous composition and processes in the Universe and deliver transformative understanding of galaxy evolution, galaxy cluster physics, and gas within the Cosmic Web. 

\pagebreak
\bibliographystyle{unsrturltrunc6}

\begin{thebibliography}{10}

\bibitem{Bergeron:1991qy}
J.~Bergeron and P.~Boiss{\'e}.
\newblock A sample of galaxies giving rise to mg {II} quasar absorption
  systems.
\newblock {\em A\&A}, 243:344--366, March 1991.

\bibitem{Tripp:1998kq}
Todd~M. Tripp, Limin Lu, and Blair~D. Savage.
\newblock The {Relationship} between {Galaxies} and {Low}-{Redshift} {Weak}
  {Ly$\alpha$} {Absorbers} in the {Directions} of {H}1821+643 and {PG}
  1116+215.
\newblock {\em ApJ}, 508:200--231, November 1998.

\bibitem{Chen:2001ys}
H.~Chen, K.~M. Lanzetta, and J.~K. Webb.
\newblock Extended c {IV} gaseous envelopes surrounding galaxies at
  z{\textless}1.
\newblock volume 240, page~37, 2001.

\bibitem{Stocke:2006yu}
John~T. Stocke, Steven~V. Penton, Charles~W. Danforth, J.~Michael Shull, Jason
  Tumlinson, and Kevin~M. McLin.
\newblock The {Galaxy} {Environment} of {O} {VI} {Absorption} {Systems}.
\newblock {\em ApJ}, 641:217--228, April 2006.

\bibitem{Prochaska:2011aa}
J.~X. {Prochaska}, D.~{Kasen}, and K.~{Rubin}.
\newblock {Simple Models of Metal-line Absorption and Emission from Cool Gas
  Outflows}.
\newblock {\em ApJ}, 734:24, June 2011.

\bibitem{Tumlinson:2017aa}
Jason Tumlinson, Molly~S. Peeples, and Jessica~K. Werk.
\newblock The {Circumgalactic} {Medium}.
\newblock {\em ARAA}, 55:389--432, August 2017.

\bibitem{McQuinn:2018aa}
M.~{McQuinn} and J.~K. {Werk}.
\newblock {Implications of the large OVI columns around low-redshift $L\_*$
  galaxies}.
\newblock {\em ApJ}, 852(33):16, 2018.

\bibitem{Faerman:2017aa}
Y.~{Faerman}, A.~{Sternberg}, and C.~F. {McKee}.
\newblock {Massive Warm/Hot Galaxy Coronae as Probed by UV/X-Ray Oxygen
  Absorption and Emission. I. Basic Model}.
\newblock {\em ApJ}, 835:52, January 2017.

\bibitem{Werk:2016aa}
J.~K. {Werk}, J.~X. {Prochaska}, S.~{Cantalupo}, A.~J. {Fox}, B.~{Oppenheimer},
  J.~{Tumlinson}, et~al.
\newblock {The COS-Halos Survey: Origins of the Highly Ionized Circumgalactic
  Medium of Star-Forming Galaxies}.
\newblock {\em ApJ}, 833:54, December 2016.

\bibitem{Oppenheimer:2016lr}
Benjamin~D. Oppenheimer, Robert~A. Crain, Joop Schaye, Alireza Rahmati,
  Alexander~J. Richings, James~W. Trayford, et~al.
\newblock Bimodality of low-redshift circumgalactic {O} vi in non-equilibrium
  eagle zoom simulations.
\newblock {\em MNRAS}, 460(2):2157--2179, August 2016.

\bibitem{Nelson:2018aa}
D.~{Nelson}, G.~{Kauffmann}, A.~{Pillepich}, S.~{Genel}, V.~{Springel},
  R.~{Pakmor}, et~al.
\newblock {The abundance, distribution, and physical nature of highly ionized
  oxygen O VI, O VII, and O VIII in IllustrisTNG}.
\newblock {\em MNRAS}, 477:450--479, June 2018.

\bibitem{Nielsen:2016aa}
Nikole~M. Nielsen, Christopher~W. Churchill, Glenn~G. Kacprzak, Michael~T.
  Murphy, and Jessica~L. Evans.
\newblock {MAGIICAT} {IV}. {Kinematics} of the {Circumgalactic} {Medium} and
  {Evidence} for {Quiescent} {Evolution} {Around} {Red} {Galaxies}.
\newblock {\em The Astrophysical Journal}, 818:171, February 2016.

\bibitem{Tumlinson:2011kx}
J.~Tumlinson, C.~Thom, J.~K. Werk, J.~X. Prochaska, T.~M. Tripp, D.~H.
  Weinberg, et~al.
\newblock The large, oxygen-rich halos of star-forming galaxies are a major
  reservoir of galactic metals.
\newblock {\em Science}, 334:948, November 2011.

\bibitem{Johnson:2015qv}
Sean~D. Johnson, Hsiao-Wen Chen, and John~S. Mulchaey.
\newblock On the possible environmental effect in distributing heavy elements
  beyond individual gaseous haloes.
\newblock {\em MNRAS}, 449:3263--3273, May 2015.

\bibitem{Werk:2014kx}
Jessica~K. Werk, J.~Xavier Prochaska, Jason Tumlinson, Molly~S. Peeples,
  Todd~M. Tripp, Andrew~J. Fox, et~al.
\newblock The {COS}-halos survey: Physical conditions and baryonic mass in the
  low-redshift circumgalactic medium.
\newblock {\em ApJ}, 792:8, September 2014.

\bibitem{Prochaska:2017aa}
J.~Xavier Prochaska, Jessica~K. Werk, G{\'a}bor Worseck, Todd~M. Tripp, Jason
  Tumlinson, Joseph~N. Burchett, et~al.
\newblock The {COS}-{Halos} {Survey}: {Metallicities} in the {Low}-redshift
  {Circumgalactic} {Medium}.
\newblock {\em ApJ}, 837:169, March 2017.

\bibitem{Johnson:2014rt}
Sean~D. Johnson, Hsiao-Wen Chen, John~S. Mulchaey, Todd~M. Tripp, J.~Xavier
  Prochaska, and Jessica~K. Werk.
\newblock Discovery of a transparent sightline at rho {$\leq$} 20 kpc from an
  interacting pair of galaxies.
\newblock {\em MNRAS}, 438:3039--3048, March 2014.

\bibitem{Burchett:2016aa}
Joseph~N. Burchett, Todd~M. Tripp, Rongmon Bordoloi, Jessica~K. Werk, J.~Xavier
  Prochaska, Jason Tumlinson, et~al.
\newblock A {Deep} {Search} {For} {Faint} {Galaxies} {Associated} {With} {Very}
  {Low}-redshift {C} {IV} {Absorbers}: {III}. {The} {Mass}- and
  {Environment}-dependent {Circumgalactic} {Medium}.
\newblock {\em ApJ}, 832(124):124, December 2016.

\bibitem{Burchett:2018aa}
J.~N. {Burchett}, T.~M. {Tripp}, Q.~D. {Wang}, C.~N.~A. {Willmer}, D.~V.
  {Bowen}, and E.~B. {Jenkins}.
\newblock {Warm-hot gas in X-ray bright galaxy clusters and the H I-deficient
  circumgalactic medium in dense environments}.
\newblock {\em MNRAS}, 475:2067--2085, April 2018.

\bibitem{Tumlinson:2013cr}
Jason Tumlinson, Christopher Thom, Jessica~K. Werk, J.~Xavier Prochaska,
  Todd~M. Tripp, Neal Katz, et~al.
\newblock The {COS-Halos} survey: Rationale, design, and a census of
  circumgalactic neutral hydrogen.
\newblock {\em ApJ}, 777:59, November 2013.

\bibitem{Bordoloi:2014lr}
Rongmon Bordoloi, Jason Tumlinson, Jessica~K. Werk, Benjamin~D. Oppenheimer,
  Molly~S. Peeples, J.~Xavier Prochaska, et~al.
\newblock The {COS}-dwarfs survey: The carbon reservoir around sub-l* galaxies.
\newblock {\em ApJ}, 796:136, December 2014.

\bibitem{Johnson:2017aa}
S.~D. {Johnson}, H.-W. {Chen}, J.~S. {Mulchaey}, J.~{Schaye}, and L.~A.
  {Straka}.
\newblock {The Extent of Chemically Enriched Gas around Star-forming Dwarf
  Galaxies}.
\newblock {\em \apjl}, 850:L10, November 2017.

\bibitem{Chen:2018aa}
H.-W. {Chen}, F.~S. {Zahedy}, S.~D. {Johnson}, R.~M. {Pierce}, Y.-H. {Huang},
  B.~J. {Weiner}, et~al.
\newblock {Characterizing circumgalactic gas around massive ellipticals at z
  {\tilde} 0.4 - I. Initial results}.
\newblock {\em \mnras}, 479:2547--2563, September 2018.

\bibitem{Smailagic:2018aa}
M.~{Smailagi{\'c}}, J.~X. {Prochaska}, J.~{Burchett}, G.~{Zhu}, and
  B.~{M{\'e}nard}.
\newblock {Extreme Circumgalactic H I and C III Absorption around the Most
  Massive, Quenched Galaxies}.
\newblock {\em \apj}, 867:106, November 2018.

\bibitem{Berg:2018aa}
T.~A.~M. {Berg}, S.~L. {Ellison}, J.~{Tumlinson}, B.~D. {Oppenheimer},
  R.~{Horton}, R.~{Bordoloi}, et~al.
\newblock {The COS-AGN survey: revealing the nature of circumgalactic gas
  around hosts of active galactic nuclei}.
\newblock {\em MNRAS}, 478:3890--3934, August 2018.

\bibitem{Yoon:2012yu}
Joo~Heon Yoon, Mary~E. Putman, Christopher Thom, Hsiao-Wen Chen, and Greg~L.
  Bryan.
\newblock Warm {Gas} in the {Virgo} {Cluster}. {I}. {Distribution} of
  {Ly$\alpha$} {Absorbers}.
\newblock {\em ApJ}, 754:84, August 2012.

\bibitem{Muzahid:2017lr}
Sowgat Muzahid, Jane Charlton, Daisuke Nagai, Joop Schaye, and Raghunathan
  Srianand.
\newblock Discovery of an {H} i-rich {Gas} {Reservoir} in the {Outskirts} of
  {SZ}-effect-selected {Clusters}.
\newblock {\em The Astrophysical Journal Letters}, 846:L8, September 2017.

\bibitem{Tejos:2012lr}
Nicolas Tejos, Simon~L. Morris, Neil H.~M. Crighton, Tom Theuns, Gabriel Altay,
  and Charles~W. Finn.
\newblock Large-scale structure in absorption: gas within and around galaxy
  voids.
\newblock {\em MNRAS}, 425:245--260, September 2012.

\bibitem{Tejos:2016qv}
Nicolas Tejos, J.~Xavier Prochaska, Neil H.~M. Crighton, Simon~L. Morris,
  Jessica~K. Werk, Tom Theuns, et~al.
\newblock Towards the statistical detection of the warm-hot intergalactic
  medium in intercluster filaments of the cosmic web.
\newblock {\em MNRAS}, 455:2662--2697, January 2016.

\bibitem{Bolcar:2017aa}
M.~R. {Bolcar}, S.~{Aloezos}, V.~T. {Bly}, C.~{Collins}, J.~{Crooke}, C.~D.
  {Dressing}, et~al.
\newblock {The Large UV/Optical/Infrared Surveyor (LUVOIR): Decadal Mission
  concept design update}.
\newblock In {\em Society of Photo-Optical Instrumentation Engineers (SPIE)
  Conference Series}, volume 10398 of {\em Society of Photo-Optical
  Instrumentation Engineers (SPIE) Conference Series}, page 1039809, September
  2017.

\bibitem{Harris:2018aa}
W.~{Harris}, K.~{France}, B.~{Fleming}, and M.~{Bolcar}.
\newblock {The LUVOIR Ultraviolet Multi-Object Spectrograph (LUMOS)}.
\newblock {\em AGU Fall Meeting Abstracts}, December 2018.

\bibitem{McNamara:2005aa}
B.~R. {McNamara}, P.~E.~J. {Nulsen}, M.~W. {Wise}, D.~A. {Rafferty},
  C.~{Carilli}, C.~L. {Sarazin}, et~al.
\newblock {The heating of gas in a galaxy cluster by X-ray cavities and
  large-scale shock fronts}.
\newblock {\em \nat}, 433:45--47, January 2005.

\bibitem{Walker:2019aa}
Stephen Walker, Aurora Simionescu, Daisuke Nagai, Nobuhiro Okabe, Dominique
  Eckert, Tony Mroczkowski, et~al.
\newblock The physics of galaxy cluster outskirts.
\newblock {\em Space Science Reviews}, 215(1):7, Jan 2019.

\bibitem{Emerick:2015aa}
A.~Emerick, G.~Bryan, and M.~E. Putman.
\newblock Warm gas in and around simulated galaxy clusters as probed by
  absorption lines.
\newblock {\em MNRAS}, 453:4051--4069, November 2015.

\bibitem{Hurier:2019aa}
G.~{Hurier}, R.~{Adam}, and U.~{Keshet}.
\newblock {First detection of a virial shock with SZ data: implication for the
  mass accretion rate of Abell 2319}.
\newblock {\em \aap}, 622:A136, February 2019.

\bibitem{Gunn:1972qy}
James~E. Gunn and J.~Richard Gott, III.
\newblock On the {Infall} of {Matter} {Into} {Clusters} of {Galaxies} and
  {Some} {Effects} on {Their} {Evolution}.
\newblock {\em ApJ}, 176:1, August 1972.

\bibitem{Fumagalli:2014aa}
Michele Fumagalli, Matteo Fossati, George K.~T. Hau, Giuseppe Gavazzi, Richard
  Bower, Ming Sun, et~al.
\newblock {MUSE} sneaks a peek at extreme ram-pressure stripping events - {I}.
  {A} kinematic study of the archetypal galaxy {ESO}137-001.
\newblock {\em MNRAS}, 445:4335--4344, December 2014.

\bibitem{Jachym2014}
P.~{J{\'a}chym}, F.~{Combes}, L.~{Cortese}, M.~{Sun}, and J.~D.~P. {Kenney}.
\newblock {Abundant Molecular Gas and Inefficient Star Formation in
  Intracluster Regions: Ram Pressure Stripped Tail of the Norma Galaxy
  ESO137-001}.
\newblock {\em \apj}, 792:11, September 2014.
\newblock \href {http://arxiv.org/abs/1403.2328} {\path{arXiv:1403.2328}},
  \href {http://dx.doi.org/10.1088/0004-637X/792/1/11}
  {\path{doi:10.1088/0004-637X/792/1/11}}.

\bibitem{Tonnesen:2007yq}
Stephanie Tonnesen, Greg~L. Bryan, and J.~H. van Gorkom.
\newblock Environmentally {Driven} {Evolution} of {Simulated} {Cluster}
  {Galaxies}.
\newblock {\em ApJ}, 671:1434--1445, December 2007.

\bibitem{Zinger:2018aa}
E.~{Zinger}, A.~{Dekel}, A.~V. {Kravtsov}, and D.~{Nagai}.
\newblock {Quenching of satellite galaxies at the outskirts of galaxy
  clusters}.
\newblock {\em MNRAS}, 475:3654--3681, April 2018.

\bibitem{Cramer2019}
W.~J. {Cramer}, J.~D.~P. {Kenney}, M.~{Sun}, H.~{Crowl}, M.~{Yagi},
  P.~{J{\'a}chym}, et~al.
\newblock {Spectacular Hubble Space Telescope Observations of the Coma Galaxy
  D100 and Star Formation in Its Ram Pressure{\ndash}stripped Tail}.
\newblock {\em \apj}, 870:63, January 2019.
\newblock \href {http://arxiv.org/abs/1811.04916} {\path{arXiv:1811.04916}},
  \href {http://dx.doi.org/10.3847/1538-4357/aaefff}
  {\path{doi:10.3847/1538-4357/aaefff}}.

\bibitem{Tremmel:2019aa}
M.~{Tremmel}, T.~R. {Quinn}, A.~{Ricarte}, A.~{Babul}, U.~{Chadayammuri},
  P.~{Natarajan}, et~al.
\newblock {Introducing ROMULUSC: a cosmological simulation of a galaxy cluster
  with an unprecedented resolution}.
\newblock {\em \mnras}, 483:3336--3362, March 2019.

\bibitem{Yoon:2017aa}
Joo~Heon Yoon and M.~E. Putman.
\newblock Ly$\alpha$ {Absorbers} and the {Coma} {Cluster}.
\newblock {\em The Astrophysical Journal}, 839:117, April 2017.

\bibitem{Wakker:2009fr}
B.~P. Wakker and B.~D. Savage.
\newblock The relationship between intergalactic h {I/O} {VI} and nearby (z
  {\textless} 0.017) galaxies.
\newblock {\em ApJS}, 182:378--467, May 2009.

\bibitem{Yoon:2013kq}
Joo~Heon Yoon and Mary~E. Putman.
\newblock The {Influence} of {Environment} on the {Circumgalactic} {Medium}.
\newblock {\em ApJL}, 772:L29, August 2013.

\bibitem{Pointon:2017aa}
S.~K. {Pointon}, N.~M. {Nielsen}, G.~G. {Kacprzak}, S.~{Muzahid}, C.~W.
  {Churchill}, and J.~C. {Charlton}.
\newblock {The Impact of the Group Environment on the O VI Circumgalactic
  Medium}.
\newblock {\em ApJ}, 844:23, July 2017.

\bibitem{Nielsen:2018aa}
N.~M. {Nielsen}, G.~G. {Kacprzak}, S.~K. {Pointon}, C.~W. {Churchill}, and
  M.~T. {Murphy}.
\newblock {MAGIICAT VI. The Mg II Intragroup Medium Is Kinematically Complex}.
\newblock {\em \apj}, 869:153, December 2018.

\bibitem{Simionescu2019}
A.~{Simionescu}, J.~{ZuHone}, I.~{Zhuravleva}, E.~{Churazov}, M.~{Gaspari},
  D.~{Nagai}, et~al.
\newblock {Constraining Gas Motions in the Intra-Cluster Medium}.
\newblock {\em \ssr}, 215:24, February 2019.
\newblock \href {http://arxiv.org/abs/1902.00024} {\path{arXiv:1902.00024}},
  \href {http://dx.doi.org/10.1007/s11214-019-0590-1}
  {\path{doi:10.1007/s11214-019-0590-1}}.

\bibitem{Mroczkowski2019}
T.~{Mroczkowski}, D.~{Nagai}, K.~{Basu}, J.~{Chluba}, J.~{Sayers}, R.~{Adam},
  et~al.
\newblock {Astrophysics with the Spatially and Spectrally Resolved
  Sunyaev-Zeldovich Effects. A Millimetre/Submillimetre Probe of the Warm and
  Hot Universe}.
\newblock {\em \ssr}, 215:17, February 2019.
\newblock \href {http://arxiv.org/abs/1811.02310} {\path{arXiv:1811.02310}},
  \href {http://dx.doi.org/10.1007/s11214-019-0581-2}
  {\path{doi:10.1007/s11214-019-0581-2}}.

\bibitem{Cai:2017aa}
Z.~{Cai}, X.~{Fan}, Y.~{Yang}, F.~{Bian}, J.~X. {Prochaska}, A.~{Zabludoff},
  et~al.
\newblock {Discovery of an Enormous Ly{$\alpha$} Nebula in a Massive Galaxy
  Overdensity at z = 2.3}.
\newblock {\em ApJ}, 837:71, March 2017.

\bibitem{Valentino:2016aa}
F.~{Valentino}, E.~{Daddi}, A.~{Finoguenov}, V.~{Strazzullo}, A.~{Le Brun},
  C.~{Vignali}, et~al.
\newblock {A Giant Ly{$\alpha$} Nebula in the Core of an X-Ray Cluster at Z =
  1.99: Implications for Early Energy Injection}.
\newblock {\em \apj}, 829:53, September 2016.

\bibitem{ODea:2004aa}
C.~P. {O'Dea}, S.~A. {Baum}, J.~{Mack}, A.~M. {Koekemoer}, and A.~{Laor}.
\newblock {Hubble Space Telescope STIS Far-Ultraviolet Observations of the
  Central Nebulae in the Cooling-Core Clusters A1795 and A2597}.
\newblock {\em \apj}, 612:131--151, September 2004.

\bibitem{Fabian:1984aa}
A.~C. {Fabian}, P.~E.~J. {Nulsen}, and K.~A. {Arnaud}.
\newblock {Diffuse Lyman-alpha emission around NGC 1275}.
\newblock {\em \mnras}, 208:179--184, May 1984.

\bibitem{Zinger2016}
E.~{Zinger}, A.~{Dekel}, Y.~{Birnboim}, A.~{Kravtsov}, and D.~{Nagai}.
\newblock {The role of penetrating gas streams in setting the dynamical state
  of galaxy clusters}.
\newblock {\em \mnras}, 461:412--432, September 2016.
\newblock \href {http://arxiv.org/abs/1510.05388} {\path{arXiv:1510.05388}},
  \href {http://dx.doi.org/10.1093/mnras/stw1283}
  {\path{doi:10.1093/mnras/stw1283}}.

\bibitem{Mandelker:2019aa}
N.~{Mandelker}, D.~{Nagai}, H.~{Aung}, A.~{Dekel}, D.~{Padnos}, and
  Y.~{Birnboim}.
\newblock {Instability of supersonic cold streams feeding Galaxies - III.
  Kelvin-Helmholtz instability in three dimensions}.
\newblock {\em \mnras}, 484:1100--1132, March 2019.

\bibitem{Voit:2015aa}
G.~M. {Voit} and M.~{Donahue}.
\newblock {Cooling Time, Freefall Time, and Precipitation in the Cores of
  ACCEPT Galaxy Clusters}.
\newblock {\em \apjl}, 799:L1, January 2015.

\bibitem{Tuttle:2008aa}
S.~E. {Tuttle}, D.~{Schiminovich}, B.~{Milliard}, R.~{Grange}, D.~C. {Martin},
  S.~{Rahman}, et~al.
\newblock {The FIREBall fiber-fed UV spectrograph}.
\newblock In {\em Ground-based and Airborne Instrumentation for Astronomy II},
  volume 7014 of {\em \procspie}, page 70141T, July 2008.

\bibitem{Lee:2016aa}
K.-G. {Lee}.
\newblock {Ly{$\alpha$} Forest Tomography of the z $>$ 2 Cosmic Web}.
\newblock In R.~{van de Weygaert}, S.~{Shandarin}, E.~{Saar}, and J.~{Einasto},
  editors, {\em The Zeldovich Universe: Genesis and Growth of the Cosmic Web},
  volume 308 of {\em IAU Symposium}, pages 360--363, October 2016.

\bibitem{Tanimura:2019aa}
H.~{Tanimura}, G.~{Hinshaw}, I.~G. {McCarthy}, L.~{Van Waerbeke}, N.~{Aghanim},
  Y.-Z. {Ma}, et~al.
\newblock {A search for warm/hot gas filaments between pairs of SDSS Luminous
  Red Galaxies}.
\newblock {\em \mnras}, 483:223--234, February 2019.

\bibitem{de-Graaff:2017aa}
A.~{de Graaff}, Y.-C. {Cai}, C.~{Heymans}, and J.~A. {Peacock}.
\newblock {Probing the missing baryons with the Sunyaev-Zel'dovich effect from
  filaments}.
\newblock {\em arXiv e-prints}, September 2017.

\bibitem{Tejos:2015lr}
Nicolas Tejos, J.~Xavier Prochaska, Neil H.~M. Crighton, Simon~L. Morris,
  Jessica~K. Werk, Tom Theuns, et~al.
\newblock Towards the statistical detection of the warm-hot intergalactic
  medium in inter-cluster filaments of the cosmic web.
\newblock {\em ArXiv e-prints}, 1506:1031, June 2015.

\bibitem{Cai:2016aa}
Z.~{Cai}, X.~{Fan}, S.~{Peirani}, F.~{Bian}, B.~{Frye}, I.~{McGreer}, et~al.
\newblock {Mapping the Most Massive Overdensity Through Hydrogen (MAMMOTH) I:
  Methodology}.
\newblock {\em \apj}, 833:135, December 2016.

\bibitem{Lee:2018aa}
K.-G. {Lee}, A.~{Krolewski}, M.~{White}, D.~{Schlegel}, P.~E. {Nugent}, J.~F.
  {Hennawi}, et~al.
\newblock {First Data Release of the COSMOS Ly{$\alpha$} Mapping and Tomography
  Observations: 3D Ly{$\alpha$} Forest Tomography at 2.05 $\lt$ z $\lt$ 2.55}.
\newblock {\em \apjs}, 237:31, August 2018.

\bibitem{Fukugita:1998vn}
M.~Fukugita, C.~J. Hogan, and P.~J.~E. Peebles.
\newblock The {Cosmic} {Baryon} {Budget}.
\newblock {\em ApJ}, 503:518--530, August 1998.

\bibitem{Dave:2001dp}
Romeel Dav{\'e}, Renyue Cen, Jeremiah~P. Ostriker, Greg~L. Bryan, Lars
  Hernquist, Neal Katz, et~al.
\newblock Baryons in the warm-hot intergalactic medium.
\newblock {\em ApJ}, 552:473--483, May 2001.

\bibitem{Bregman2007}
J.~N. {Bregman}.
\newblock {The Search for the Missing Baryons at Low Redshift}.
\newblock {\em \araa}, 45:221--259, September 2007.
\newblock \href {http://arxiv.org/abs/0706.1787} {\path{arXiv:0706.1787}},
  \href {http://dx.doi.org/10.1146/annurev.astro.45.051806.110619}
  {\path{doi:10.1146/annurev.astro.45.051806.110619}}.

\bibitem{Tepper-Garcia:2013aa}
T.~{Tepper-Garc{\'{\i}}a}, P.~{Richter}, and J.~{Schaye}.
\newblock {Absorption signatures of warm-hot gas at low redshift: Ne VIII}.
\newblock {\em \mnras}, 436:2063--2081, December 2013.

\bibitem{Corlies:2016aa}
Lauren Corlies and David Schiminovich.
\newblock Empirically {Constrained} {Predictions} for {Metal}-line {Emission}
  from the {Circumgalactic} {Medium}.
\newblock {\em ApJ}, 827:148, August 2016.

\bibitem{Corlies:2018aa}
L.~{Corlies}, M.~S. {Peeples}, J.~{Tumlinson}, B.~W. {O'Shea}, N.~{Lehner},
  J.~C. {Howk}, et~al.
\newblock {Figuring Out Gas {\amp} Galaxies in Enzo (FOGGIE). II. Emission from
  the z=3 Circumgalactic Medium}.
\newblock {\em arXiv e-prints}, November 2018.

\end{thebibliography}

\end{document}